\documentstyle[12pt,epsf,psfig,epsfig,graphics,amssymb,here]{article}

\def\be{\begin{equation}}
\def\ee{\end{equation}}
                         \def\bearr{\begin{eqnarray}}
                         \def\eearr{\end{eqnarray}}
\def\benum{\begin{enumerate}}
\def\eenum{\end{enumerate}}
\def\bitem{\begin{itemize}}
\def\eitem{\end{itemize}}

\def\beq{\begin{equation}}
\def\eeq{\end{equation}}
\def\bea{\begin{eqnarray}}
\def\beaa{\begin{eqnarray*}}
\def\eea{\end{eqnarray}}
\def\eeaa{\end{eqnarray*}}
\def\bq{\begin{quote}}
\def\eq{\end{quote}}
\def\gappeq{\mathrel{\rlap {\raise.5ex\hbox{$>$}}
{\lower.5ex\hbox{$\sim$}}}}

\def\lappeq{\mathrel{\rlap{\raise.5ex\hbox{$<$}}
{\lower.5ex\hbox{$\sim$}}}}

\begin{document}

\begin{center}
{\it Submitted to the European Particle Physics Strategy Preparatory Group}
\end{center}


\vspace{0.5cm}

\begin{center}
{\Large \bf 
SAPPHiRE: a Small $\gamma \gamma$ Higgs Factory}\\
\vspace*{1.0cm}
{\bf S.~A.~Bogacz}$^1$,
{\bf J.~Ellis}$^{2,3}$,
{\bf L.~Lusito}$^4$,
{\bf D.~Schulte}$^3$,
{\bf T.~Takahashi}$^5$,
{\bf M.~Velasco}$^4$,
{\bf M.~Zanetti}$^6$
and  
{\bf F.~Zimmermann$^3$}\\
\vspace{1.0cm}
{\small 
$^1$ Thomas Jefferson National Accelerator Facility, Newport News, VA 23606, USA \\
$^2$ Theoretical Particle Physics and Cosmology Group, \\
Physics Department, King's College London, London WC2R 2LS, UK\\
$^3$ CERN, CH-1211 Geneva 23, Switzerland  \\
$^4$ Physics Department, Northwestern University, Evanston, IL 60201, USA  \\
$^5$ Physics Department, Hiroshima University, 1-3-1 Kagamiyama, Higashi-Hiroshima 739-8526 Japan \\
$^6$ Laboratory for Nuclear Science, MIT, Cambridge, MA 02139, USA \\
}
\vspace{1.0cm}
{\bf Abstract} \\
\end{center}

A new particle with mass $\sim 125$~GeV that resembles the Higgs
boson has recently been discovered  by ATLAS and CMS. We propose a low-energy $\gamma\gamma$ collider as a
cost- and time-efficient option for a Higgs factory capable of studying this particle in detail.
In the past, this option has been suggested as a possible application
of the CLIC two-beam accelerator technology (the CLIC Higgs Experiment, CLICHE)
or as an option for the ILC. Here we propose a design based on a pair of
$\sim 10$~GeV recirculating Linacs   
(Small Accelerator for Photon-Photon Higgs production using Recirculating Electrons, SAPPHiRE) 
similar in design to those proposed for the LHeC.   
We present parameters for the $e^-$ beams and sketch a laser backscattering system capable of producing
a $\gamma\gamma$ peak luminosity of $0.36 \times 
10^{34}$~cm$^{-2}$s$^{-1}$ with $E_{CM}(\gamma \gamma) \sim 125$~GeV.
A $\gamma \gamma$ collider with such a luminosity could be
used to measure accurately the mass, $\bar b b$, $WW^\ast$, and $\gamma \gamma$
decays of the Higgs boson. We also comment on possible synergies with other projects
such as LHeC, the ILC or CLIC, 
and on other physics prospects in $\gamma\gamma$ and $e^- \gamma$ collisions.\\

\newpage
\section{Introduction}

The discovery by ATLAS and CMS~\cite{:2012gu} of a new boson $h$ with mass $\sim 125$ to 126 GeV 
that decays into photon pairs and $Z$ pairs naturally focuses attention on possible future 
accelerator facilities that would offer timely and cost-effective opportunities for future detailed
studies of its properties. The LHC has already made rapid strides towards the identification 
of the $h$ particle, demonstrating that it cannot have unit spin and that its couplings to other 
particles are approximately proportional to their masses, as would be expected for a Higgs boson. 
The LHC will make much more progress in the coming years, determining its spin and measuring 
its couplings much more accurately, with the possibility of measuring its trilinear self-coupling
with the high-luminosity upgrade of the LHC.

However, proton-proton collisions are not ideal for studying the properties of a Higgs boson, and now
is the time to discuss in the most open-minded way options for supplementing the studies possible with the LHC. 
For example, electron-positron and muon-antimuon collisions offer much cleaner experimental collisions, 
and so are attracting growing attention. Here we advocate another option for detailed studies
of the $h$ boson, namely $\gamma  \gamma$ collisions. These are often regarded as adjuncts 
to linear $e^+ e^-$ colliders such as the ILC or CLIC~\cite{CLIC}. However, some years ago we proposed as an alternative
equipping the low-energy CLIC demonstrator project CLIC-1, which is designed to collide two
low-emittance $e^-$ beams with energies $\sim 70$~GeV, with a laser backscattering
system capable of producing high-luminosity $\gamma \gamma$ collisions with
$E_{CM}(\gamma \gamma) \sim 120$~GeV (the CLIC Higgs Experiment, CLICHE)~\cite{asner}.

Here we propose an alternative concept based on a pair of
$\sim 10$~GeV recirculating Linacs similar in design to those proposed for
the LHeC~\cite{lheccdr} (Small Accelerator for Photon-Photon Higgs production using Recirculating Electrons, 
SAPPHiRE). 
Equipped with with a laser backscattering
system similar to that proposed for CLICHE, this facility should be capable of a 
$\gamma\gamma$ peak luminosity of $0.36 \times 
10^{34}$~cm$^{-2}$s$^{-1}$, sufficient to produce tens of thousands of $h$ particles
per year in very clean experimental conditions.
Obvious advantages of a $\gamma\gamma$-based Higgs factory include (1) the 
lower beam energy required to produce the Higgs boson in the $s$ channel, namely about 80~GeV,
as compared with 120~GeV required in $e^{+}e^{-}$ collisions - allowing for
efficient recirculation and for an RF installation that is about 10 times smaller and hence a significant saving, 
(2) the possibility of high polarization in both the primary $e^{-}$ and the 
colliding $\gamma$ beams - in contrast to the  case of $e^{+}$, and 
(3) the absence of the need to produce positrons - another potential large saving and simplification. 

In subsequent sections of this paper, we first recall some aspects of the accelerator 
requirements for $\gamma \gamma$ Higgs factories, including the
$e^-$ beam parameters and the laser backscattering system, that would be needed to
attain a luminosity sufficient to study Higgs physics. We then review the CLICHE
option~\cite{asner} before introducing SAPPHiRE and discussing it in more detail. We then review
briefly some of the most interesting physics measurements possible with
a $\gamma \gamma$ Higgs Factory, which could include accurate measurements of $M_h$, $\Gamma(h
\rightarrow \gamma \gamma)\times {\cal BR}(h\rightarrow {\bar b}b)$,
$\Gamma(h \rightarrow \gamma \gamma)\times {\cal BR}(h\rightarrow WW^\ast)$,
$\Gamma(H \rightarrow \gamma \gamma)\times {\cal BR}(H\rightarrow\gamma\gamma)$, 
and the CP properties of the $ h \rightarrow \gamma\gamma$ coupling, following~\cite{asner}. 
We conclude by discussing comparisons and possible synergies with other accelerator projects.

\section{Concepts for a $\gamma \gamma$ Higgs Factory}

\subsection{CLICHE}

\begin{figure}
\begin{center}
\epsfbox{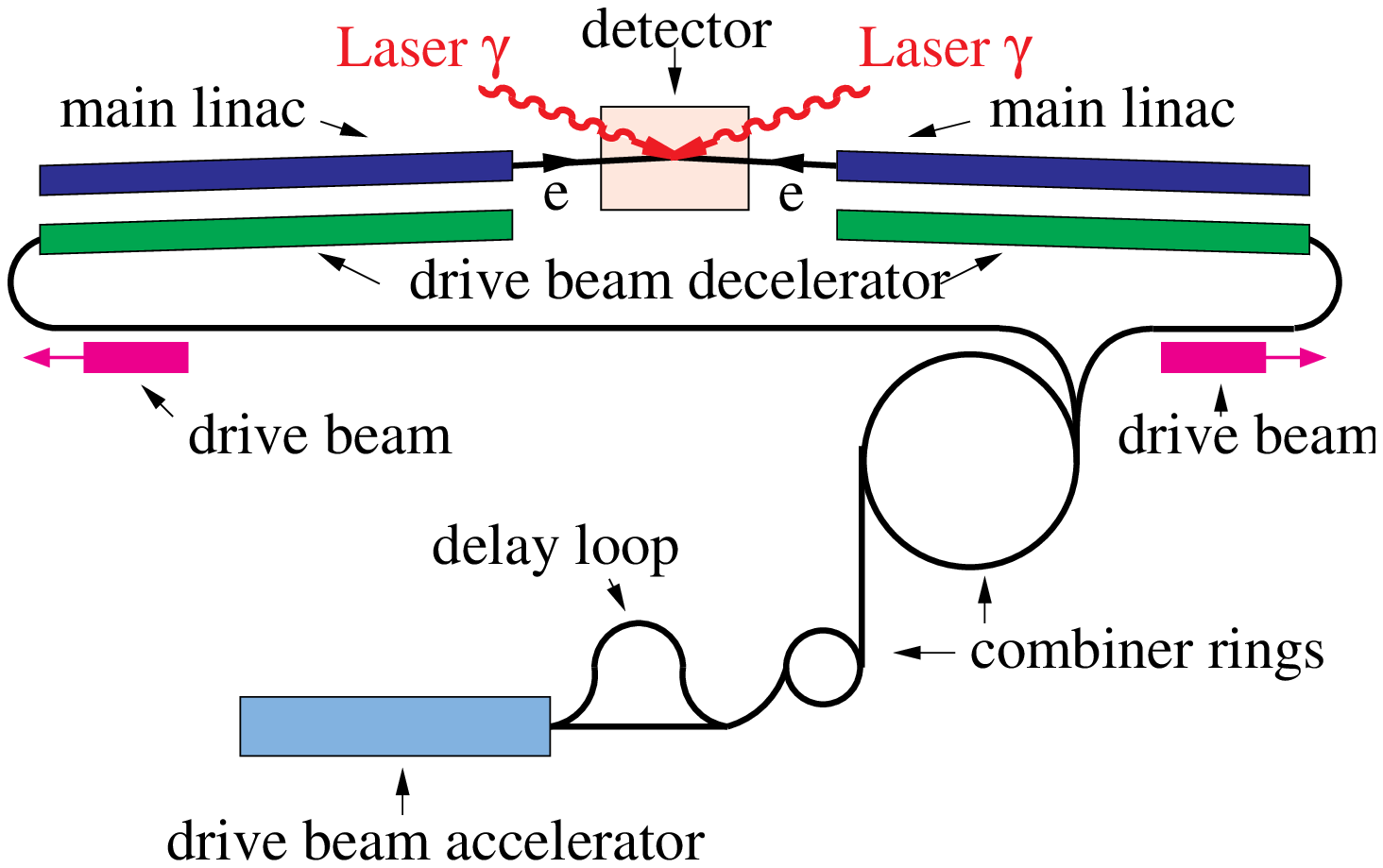}
\end{center}
\caption{\it Sketch of the possible layout of a $\gamma\gamma$ collider based on CLIC~1, the CLICHE concept~\cite{asner}.}
\label{fig:scheme}
\end{figure}

\begin{figure}[t]
\begin{center}
\mbox{\epsfig{file=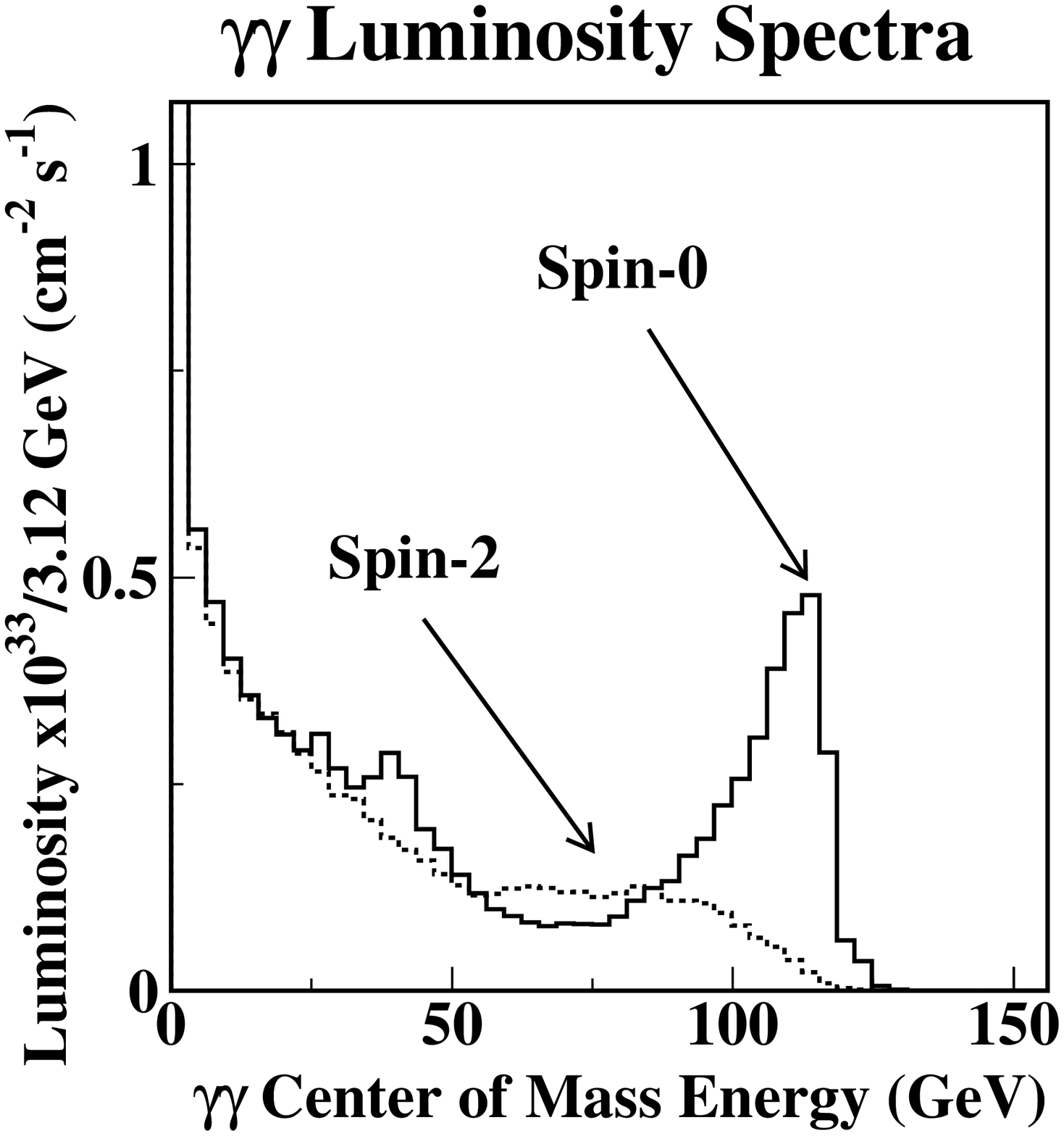,height=8cm}}
\mbox{\epsfig{file=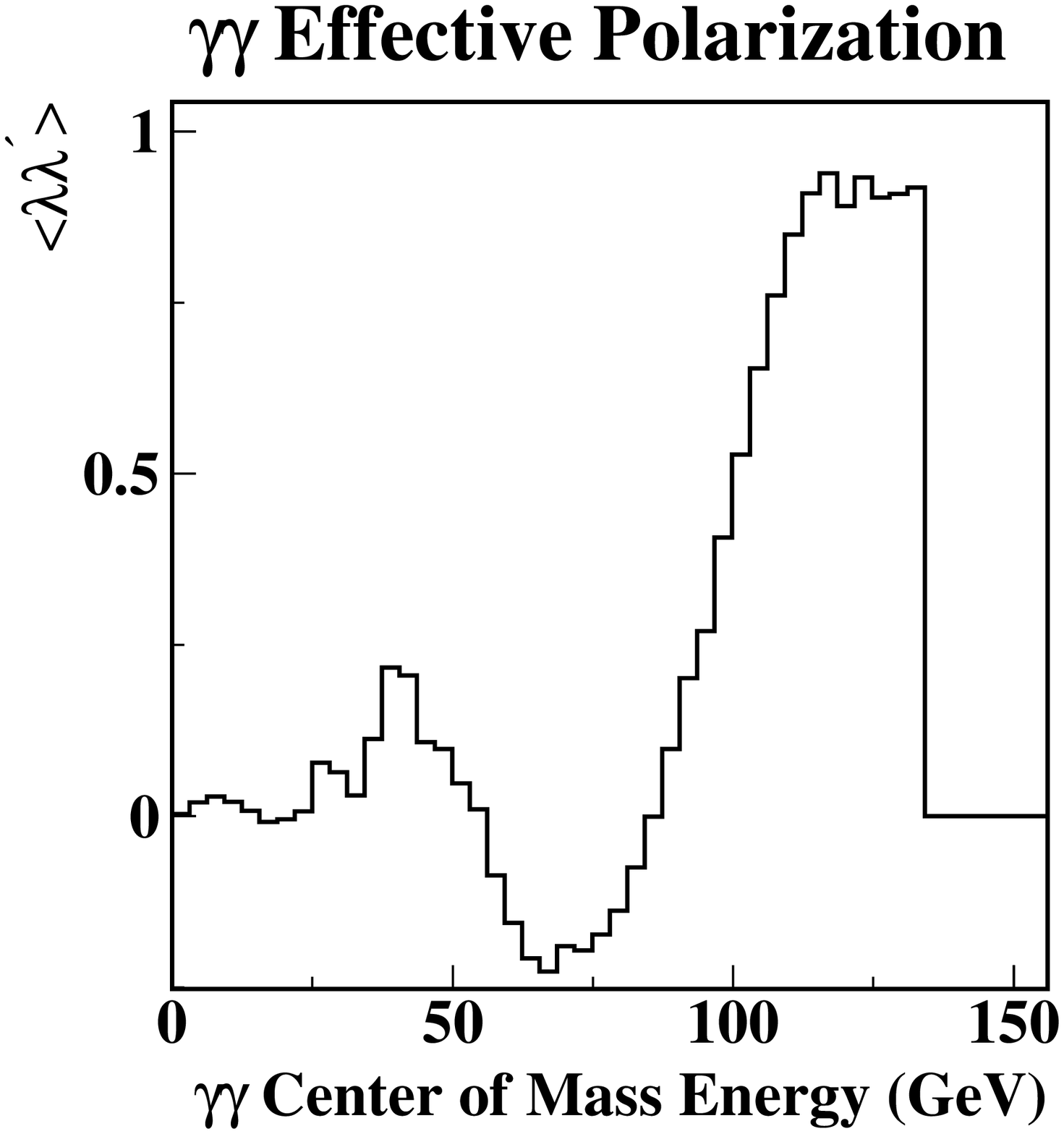,height=8cm}}
\end{center}
\caption[.]{\label{fig:spectra}\it
Luminosity spectra and beam polarization as functions of
$E_{CM}$($\gamma\gamma$) for the CLICHE parameters~\cite{asner} for 75~GeV electrons
obtained with {\tt DIMAD}~\cite{dimad} and {\tt CAIN}~\cite{cain2} for 
 ${{\cal L}_{ee}=4.8 \times10^{34} \, {\rm cm}^{-2}s^{-1}}$.
}
\end{figure}

The CLIC design for a high-energy, high-luminosity electron-positron 
linear collider uses a two-beam acceleration scheme, with conventional
normal-conducting structures accelerating the low-energy, high-intensity drive beams.
The main beams are accelerated by RF structures whose power is obtained by decelerating
the drive beams in parallel beam lines. Two-beam acceleration has been demonstrated 
successfully in CLIC test facilities (CTF1, CTF2 and CTF3). A possible future step, called CLIC~1, 
would provide a full-scale test of beam dynamics and power handling using a 70-GeV CLIC module.
A pair of such modules could provide a large geometric luminosity for $e^- e^-$ collisions.
It was proposed in~\cite{asner} that suitable laser backscattering systems could a provide a large
effective luminosity for $\gamma \gamma$ collisions with centre-of-mass energies around the
mass of a light Higgs boson, providing a relatively economical $\gamma \gamma$ Higgs factory,
the CLIC Higgs Experiment (CLICHE). Fig.~\ref{fig:scheme} displays a sketch of the possible layout of CLICHE,
and Table~\ref{t:gamma} lists some example parameters for CLICHE, as optimized in~\cite{asner} for a
hypothetical Higgs mass $M_h \sim 115$~GeV. These parameters should be adjusted for the current
$M_h \sim 125$~GeV: in particular, the beam energy should be increased by $\sim 5$~GeV, and the
electric power required would increase correspondingly. Table~\ref{t:gamma2} lists some example
parameters for the mercury laser system and photon beams of CLICHE: these parameters should also be reviewed,
in light of technological developments since~\cite{asner} and the increased centre-of-mass energy.
The CLICHE $\gamma \gamma$ luminosity spectra for the spin-0 and spin-2 states are shown in the
left panel of Fig.~\ref{fig:spectra}, and the effective polarization is shown in the right panel.

\begin{table}[htbp]
\caption{\it Example parameters for $\gamma \gamma$ colliders based on CLIC-1 (CLICHE, left column), as optimized for $M_h \sim 115$~GeV~\cite{asner}, and a pair of recirculating superconducting 
linacs (SAPPHiRE, right column) optimized for $M_h \sim 125$~GeV.}
\label{t:gamma}
\begin{center}
\begin{tabular}{lc||c|c}
\hline
Variable & Symbol & CLICHE~\cite{asner} & SAPPHiRE \\
\hline\hline
Total electric power &
$P$ & 150~MW & 100~MW \\
Beam energy & $E$ & 75~GeV & 80~GeV \\
Beam polarization & $P_e$ &0.80 & 0.80 \\
Bunch population & $N$ & $4\times 10^{9}$ & $10^{10}$ \\
Number of bunches per train & $n_{b}$ & 154 &  --- \\
Number of trains per rf pulse & $n_{t}$ & $11$ &  --- \\
Repetition rate & $f_{\rm rep}$ & 100 Hz & cw \\
Average bunch frequency & $\left< f_{\rm bunch}\right>$ & 169 kHz & 200 kHz \\ 
Average beam current & $I_{\rm beam}$ &  0.11 mA & 0.32 mA \\
RMS bunch length & $\sigma_{z}$ & 30 $\mu$m & 30 $\mu$m \\
Crossing angle & $\theta_{c}$ & $\ge 20$ mrad & $\ge 20$ mrad \\
Normalised horizontal emittance & 
$\epsilon_{x}$ & 1.4\,$\mu$m & 5\,$\mu$m \\
Normalised vertical emittance & 
$\epsilon_{y}$ & 0.05\,$\mu$m & 0.5\,$\mu$m \\
Nominal horizontal beta function at the IP &
$\beta_{x}^{\ast}$ & 2\,mm & 5\,mm \\
Nominal vertical beta function at the IP &
$\beta_{y}^{\ast}$ & 20\,$\mu$m & 0.1\,mm \\
Nominal RMS horizontal IP spot size &
$\sigma_{x}^{\ast}$ &  138 \, nm & 400\,nm \\
Nominal RMS vertical IP spot size &
$\sigma_{y}^{\ast}$ & 2.6\, nm & 18\,nm \\
Nominal RMS horizontal CP spot size &
$\sigma_{x}^{C, \ast}$ & 154\, nm & 400\,nm \\
Nominal RMS vertical CP spot size &
$\sigma_{y}^{C, \ast}$ & 131\, nm  & 180\,nm \\
e$^{-}$e$^{-}$ geometric luminosity & ${\cal L}$ &
$4.8 \times 10^{34}$~cm$^{-2}$s$^{-1}$ & $2.2 \times 10^{34}$~cm$^{-2}$s$^{-1}$ \\
\hline
\end{tabular}
\end{center}
\end{table}

\begin{table}[htbp]
\caption{\it Example parameters for the CLICHE mercury laser system~\cite{asner},
and for the SAPPHiRE laser system, assuming 
${{\cal L}_{ee}=4.8 \times 10^{34} \, \rm cm^{-2}s^{-1}}$ and
${{\cal L}_{ee}=2.2 \times 10^{34}\, \rm cm^{-2}s^{-1}}$, respectively. 
}
\label{t:gamma2}
\begin{center}
\begin{tabular}{lc||c|c}
\hline
Variable & Symbol & CLICHE~\cite{asner} & SAPPHiRE \\
\hline\hline
{Laser beam parameters} & & & \\
\hline
  Wavelength        & $\lambda_L$ & 0.351  $\mu$m  & 0.351  $\mu$m \\
  Photon energy & $\hbar\omega_L$ & 3.53  eV = 5.65$\times 10^{-19}$ J & 3.53  eV \\
  Number of laser pulses per second & $N_L$ & 169400\,s$^{-1}$  & 200000\,s$^{-1}$\\
  Laser peak power    & $W_L$ & 2.96$\times 10^{22}$ W/m$^2$  & 6.3$\times 10^{21}$ W/m$^2$\\
  Laser peak photon density    &       & 5.24$\times 10^{40}$ photons/m$^2$/s & 1.1$\times 10^{40}$ photons/m$^2$/s\\
\hline
{Photon beam} & &  & \\
\hline
Number of photons per electron bunch & $N_{\gamma}$ & $9.6\times10^{9}$ & $1.2\times10^{10}$\\
$\gamma\gamma$ luminosity for $E_{\gamma\gamma}\ge 0.6 E_{CM}$ & ${\cal L}_{\gamma\gamma}^{peak}$ & $3.6\times10^{33}$ cm$^{-2}$s$^{-1}$ & $3.6\times10^{33}$ cm$^{-2}$s$^{-1}$ \\
\hline
\end{tabular}
\end{center}
\end{table}

\subsection{SAPPHiRE}

\begin{figure}
\begin{center}
\hspace{-5cm}
\vspace{3cm}
\includegraphics[totalheight=0.45\textheight, angle=270]{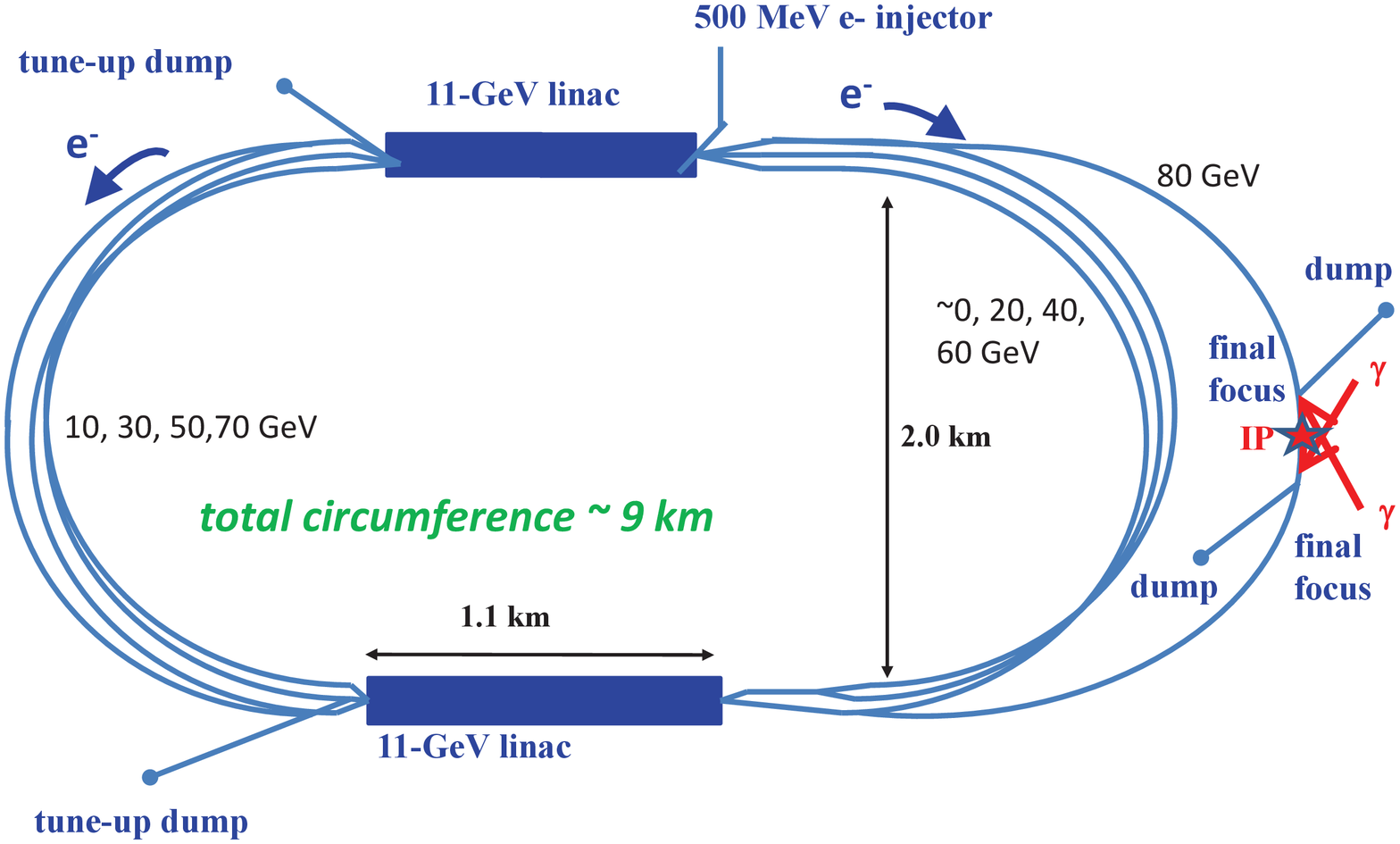}
\end{center}
\caption{\it Sketch of a layout for a $\gamma\gamma$
collider based on recirculating superconducting linacs -- the SAPPHiRE concept.}
\label{fig:LHeC}
\end{figure}

Here we propose an alternative strategy for realizing a $\gamma \gamma$ Higgs factory,
inspired by the recent design of the Large Hadron-electron Collider (LHeC)~\cite{lheccdr} that employs
a pair of recirculating linacs capable of increasing the $e^-$ energy by $\sim 10$~GeV
in each pass. A $\gamma \gamma$ Higgs factory would require an $e^- e^-$ centre-of-mass
energy of $\sim 125$~GeV/0.8/2 $\sim 80$~GeV. 
This would be achieved in SAPPHiRE via four passes through two superconducting   
recirculating linacs, as illustrated in Fig.~\ref{fig:LHeC}. Compared to the LHeC,
one additional arc is required on either side, corresponding to beam energies of 70 and 80 GeV, 
respectively. The 80~GeV arc is split into two halves with the collision point at the centre.   

The last column of Table~\ref{t:gamma} compiles a list of example parameters for SAPPHiRE, 
which would meet the luminosity target of ${\cal L}_{\gamma \gamma} \sim 2 \times 10^{34}$~cm$^{-2}$s$^{-1}$.
As in the case of CLICHE, the photon beams are produced by Compton backscattering of laser light off the 80~GeV 
electron bunches, and the Compton scattering points are taken to be about 1~mm away from the main $\gamma \gamma$ collision point.
Laser pulses are required at a rate of 200 kHz. For a photon energy of 3.53~eV, 
equivalent to a laser wavelength of 351~nm, we have $x = 4.3$ (where 
$E_{\gamma,{\rm max}} = x/(1+x) E \approx $ 65~GeV).   
These laser parameters are similar to those proposed for
the CLICHE design, as listed in Table~\ref{t:gamma2}~\footnote{As an alternative to the mercury laser technology
considered in~\cite{asner}, nowadays it it conceivable also that
fibre lasers could provide the performance required (see e.g.~\cite{laseripac12}).}.
The laser parameters can be relaxed by pulse stacking in an optical resonant cavity, e.g.~with ten recirculations
between collisions (implying a total path length of 150\,m in the optical cavity)  
and with a single-pass power reflectivity of $R=99.99\%$, the power enhancement  
is a factor of 100.  
The electron beam cross section at the Compton conversion point
is about 4 times larger than for CLICHE. For efficient conversion, the total energy of a laser pulse should be a few Joule, 
e.g., 1~TW peak power and 5~ps pulse length, implying 1~MW average power.

\begin{figure}[htbt]
\begin{center}
\mbox{\epsfig{file=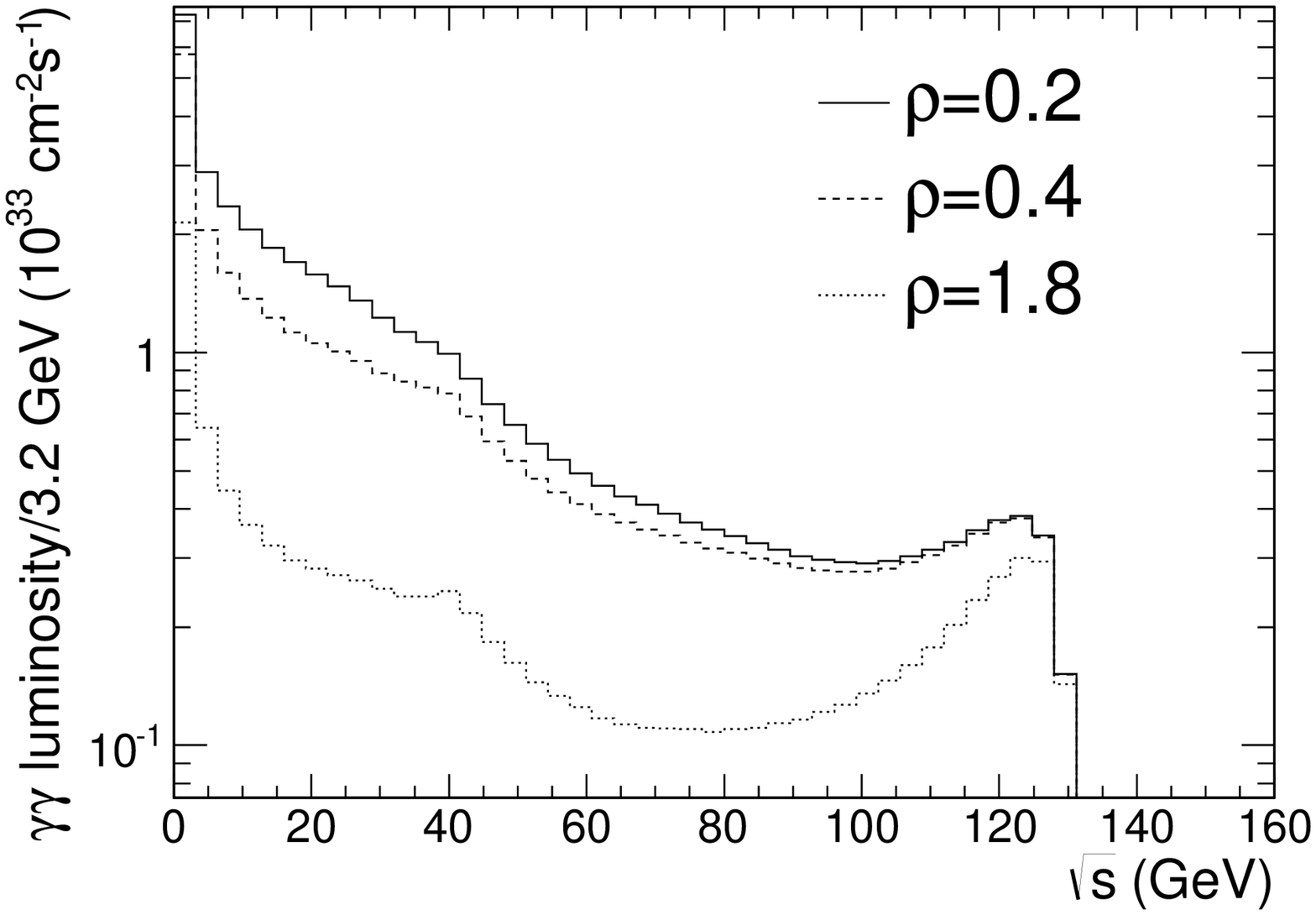,height=5.3cm}}
\mbox{\epsfig{file=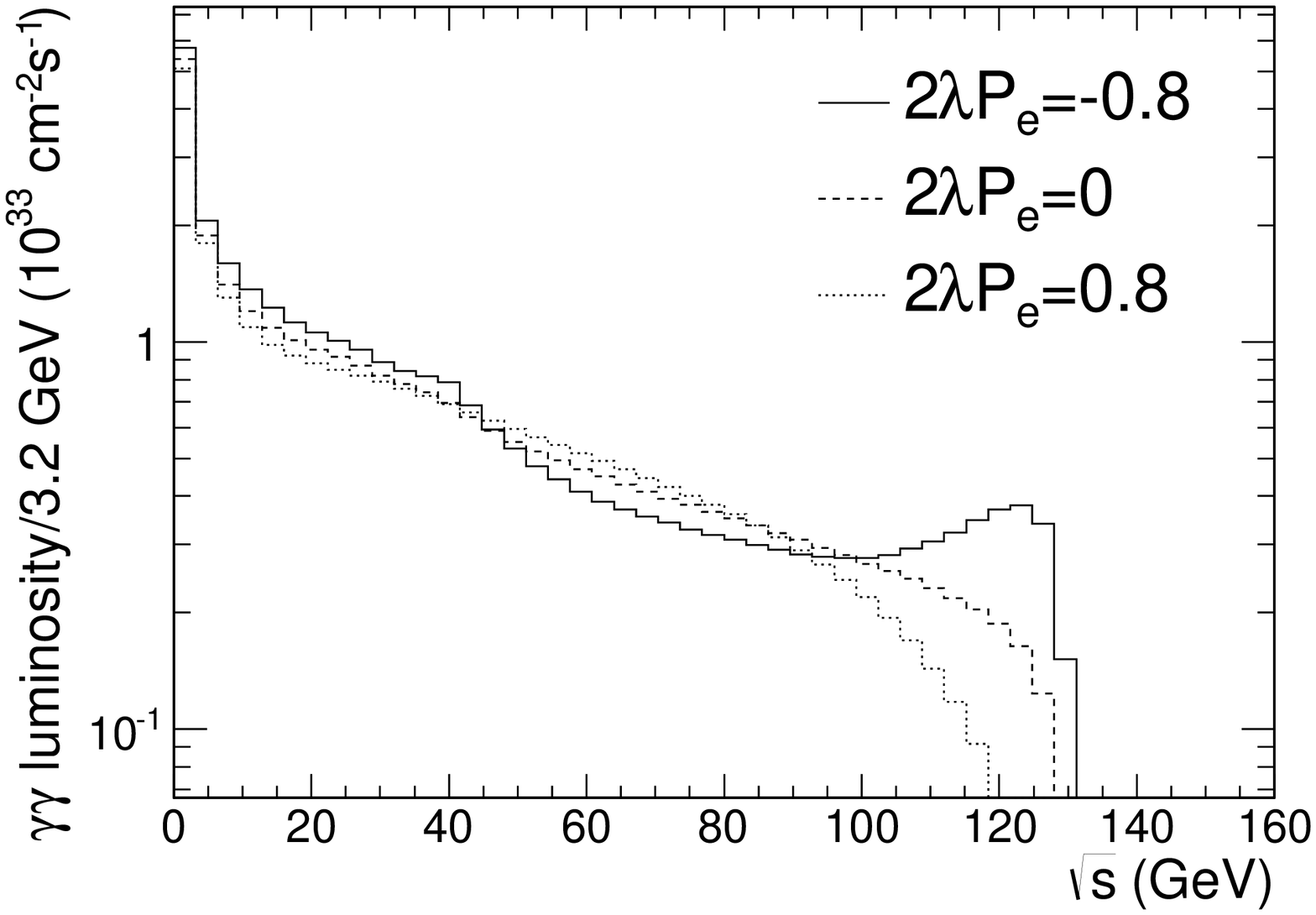,height=5.3cm}}
\end{center}
\caption[.]{\label{fig:sapphireSpectra}\it
Luminosity spectra for SAPPHiRE as functions of $E_{CM}$($\gamma\gamma$), computed using {\tt Guinea-Pig}~\cite{GP}
for three possible normalized distances (left) and different polarizations of incoming particles (right).
}
\end{figure}

The SAPPHiRE $\gamma \gamma$ luminosity spectra are shown in the plots of Fig.~\ref{fig:sapphireSpectra}.
In the left panel various normalized distances $\rho \equiv l_{\rm CP-IP}/(\gamma \sigma_{y}^{\ast})$, 
(where $l_{\rm CP-IP}$ is the distance between  the IP and the Compton backscattering point, 
$\gamma$ is the electron boost factor and $\sigma_{y}^{\ast}$ is the vertical electron beam width at the IP) 
are compared, with $\rho=0.4$ corresponding to the SAPPHiRE parameters listed in
Table~\ref{t:gamma}.
The right plot shows the effect of different beam and laser polarization configurations (with $P_e$   the polarization
of the electrons and $\lambda$ the polarization of the laser photons). 
The SAPPHiRE electron beams polarization is assumed to be 80\%.

The energy loss per arc is given by
\begin{equation} 
E_{\rm arc} {\rm [GeV]} =8.846\times10^{-5}\;   \frac{(E\, {\rm [GeV]})^4}{ 2\rho\; {\rm [m]}}\; .
\end{equation}
For a bending radius of $\rho=764$~m, as in the LHeC design, the energy loss in the various arcs is 
summarized in Table~\ref{eloss}. We see that each beam loses
about 5 GeV in energy, which can be compensated by increasing 
the voltages of the two linacs from 10~GV to 10.63~GV.
The largest energy loss due to synchrotron radiation for beams in a common arc occurs at 70~GeV,
and amounts to 1.39~GeV, or 2\%. With a dispersion of 0.1\,m (see~\cite{bogacz}),
the orbit change would be 2\,mm. The two beams would certainly fit into a common beam pipe.

\begin{table}[htbp]
\caption{\it Energy losses and energy spread induced in the 8 arcs of SAPPHiRE.}
\label{eloss}
\begin{center}
\begin{tabular}{c|c|c}
\hline
beam energy [GeV] & $\Delta E_{\rm arc}$  [GeV] & $\Delta \sigma_{E}$ [MeV]\\
\hline
10 & 0.0006 & 0.038 \\
20 & 0.009 & 0.43 \\
30 & 0.05 & 1.7 \\
40 & 0.15 & 5.0 \\
50 & 0.36 & 10 \\
60 & 0.75 & 20 \\
70 & 1.39 & 35 \\
80 (1/2 arc) & 1.19 & 27 \\
\hline
total & 3.89 & 57 \\ 
\hline
\end{tabular}
\end{center}
\end{table}

The additional energy spread from synchrotron radiation is given by 
\begin{equation}
\Delta \sigma_{E}^{2} = \frac{55 \alpha (\hbar c)^{2}}{48 \sqrt{3}} \gamma^{7} \frac{\pi}{\rho^{2}}\; ,   
\end{equation}
where $R\approx 1$~km is the geometric radius, and $\rho$ is the dipole 
bending radius in the arc. 
The total rms energy spread induced by synchrotron radiation is only 0.071\%,
as also listed Table~\ref{eloss}.

The emittance growth due to synchrotron radiation is given by 
\begin{equation}
\Delta \epsilon_{/N} = \frac{2 \pi}{3} \frac{c_{q} r_{e}}{\rho^{2}} \gamma^{6} \left< H \right>\; ,
\end{equation}
where $C_q =3.8319\times 10^{-13}$~m, and $\rho$ is the bending radius. 
For the LHeC design with $l_{\rm bend}\approx 40$~m the total length of the bending magnets 
per optical (TME) cell, and $\rho=764$~m, 
we find $\left< H \right> =1.2\times 10^{-3}$~m \cite{bogacz}, 
which is close to the ``useful and realistic'' minimum emittance optics given in~\cite{teng}. 
At 60 GeV the emittance growth of the LHeC optics design is 13 microns, too high for our purpose, 
and the extrapolation to 80 GeV with the sixth power of the energy is unfavourable. 
However, \cite{teng} also gives the scaling law 
$\left< H \right> \propto l_{\rm bend}^{3}/\rho^{2}$.  
This suggests that by reducing the cell length and associated dipole length by a factor of 4 
(to a total length of $l_{\rm bend}=10$~m per cell)
we can reduce the horizontal normalized emittance growth at 80 GeV to 1 micron,
which would be adequate for SAPPHiRE.

Beams with an emittance ratio of 10 can be produced with a flat-beam electron gun using the transformer concept
described in~\cite{brinkmann}.
Starting with a normalized uncorrelated emittance of 4-5\,$\mu$m and a bunch charge of 0.5\,nC, 
the injector test facility at the Fermilab A0 line achieved emittances of 40\,$\mu$m horizontally 
and 0.4\,$\mu$m vertically, with an emittance ratio of 100~\cite{piot}. 
For the $\gamma \gamma$ collider we need a similar emittance ratio of 10, 
but a bunch charge three times larger (1.6\,nC) and 
a smaller initial emittance of ~1.5\,$\mu$m. 
These parameters are within the present state of the art (e.g., the LCLS photoinjector routinely achieves 
1.2~$\mu$m emittance with a 1\,nC  bunch charge).

We conclude that there are no obvious showstoppers for the SAPPHiRE concept, which employs
plausible accelerator parameters.

\section{Physics with a $\gamma \gamma$ Higgs Factory}

We base our short discussion here on the
exploratory studies of the CLICHE $\gamma \gamma$ collider, incorporating modifications
associated with the measured Higgs mass $M_h \sim 125$~GeV compared to the
hypothetical value of 115~GeV assumed in~\cite{asner}.
As discussed in some detail there, several important measurements of Higgs properties can be
made at a Higgs factory and running around the Higgs threshold
offers important advantages for several analyses. Ref.~\cite{asner} also discusses
other physics opportunities that we do not dwell upon here.

\begin{figure}[h]
\begin{center}
\resizebox{\textwidth}{!}
{\epsfig{file=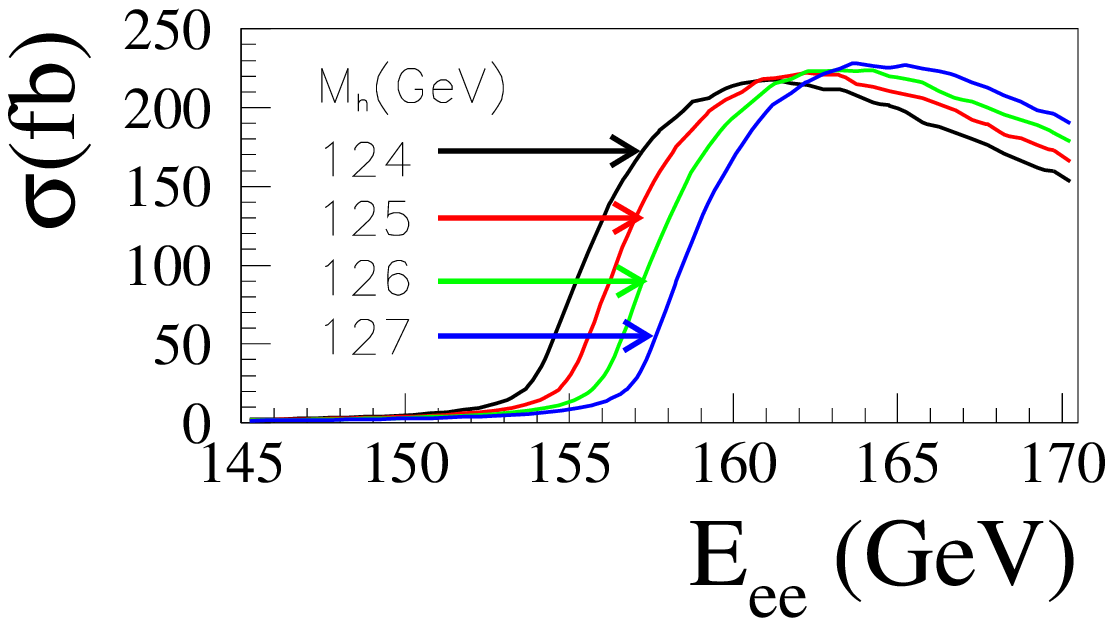,height=8.5cm}
\epsfig{file=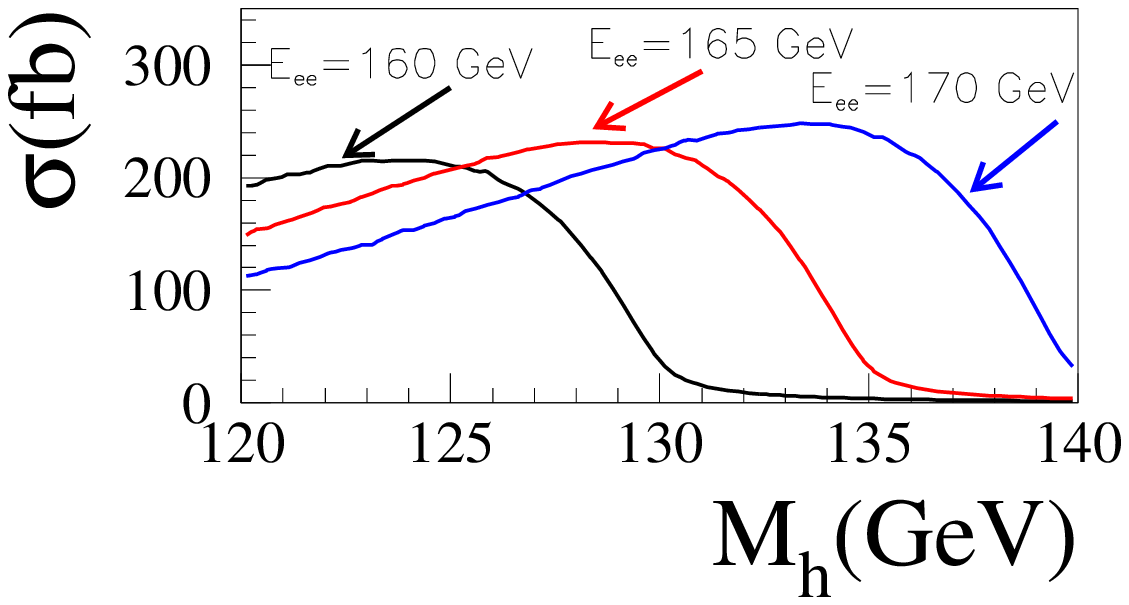,height=8.5cm}}
\end{center}
\caption[.]{\label{fig:excitation}\em
(a) The cross sections for $\gamma \gamma \rightarrow h$ for 
different values of $M_h$ as functions of $E_{CM} (e^- 
e^-)$. (b) The cross section for $\gamma \gamma \rightarrow h$
as a function of $M_h$ for three different values of  $E_{CM} (e^- e^-)$.
We assume that the electrons are have 80\% longitudinal polarization
and that the lasers are circularly polarized, so that the produced 
photons are highly circularly polarized at their maximum energy. 
}
\end{figure}

\noindent
\underline{Higgs Production}\\
The excitation curves for Higgs boson production are shown in
Fig.~\ref{fig:excitation} assuming 80\% longitudinal polarization for the electron beams
and circularly polarized lasers.  As seen in the left panel, 
the Higgs production cross section rises rapidly for 154~GeV $< E_{CM} (e^- e^-) < 164$~GeV,
providing a physics opportunity for a $\gamma \gamma$ collider obtained by laser
backscattering from a pair of $e^-$ beams with energies $\sim 80$~GeV. 
The right panel of Fig.~\ref{fig:excitation} shows the cross section as a function of Higgs mass
in the range 120~GeV $< M_h < 140$~GeV for three
choices of $E_{CM} (e^- e^-)$. The excitation curve is decreased by a factor $\sim 3$ if the
electron beams are unpolarized. The nominal luminosity of CLICHE or SAPPHiRE would
yield $\sim 20,000$ Higgs bosons per year: increasing the SAPPHiRE beam power by a
factor $\sim 2$ would increase the luminosity and hence the number of Higgs bosons produced by $\sim 4$.

\begin{figure}[htbp]
\begin{center}
\resizebox{\textwidth}{!}
{\epsfig{file= 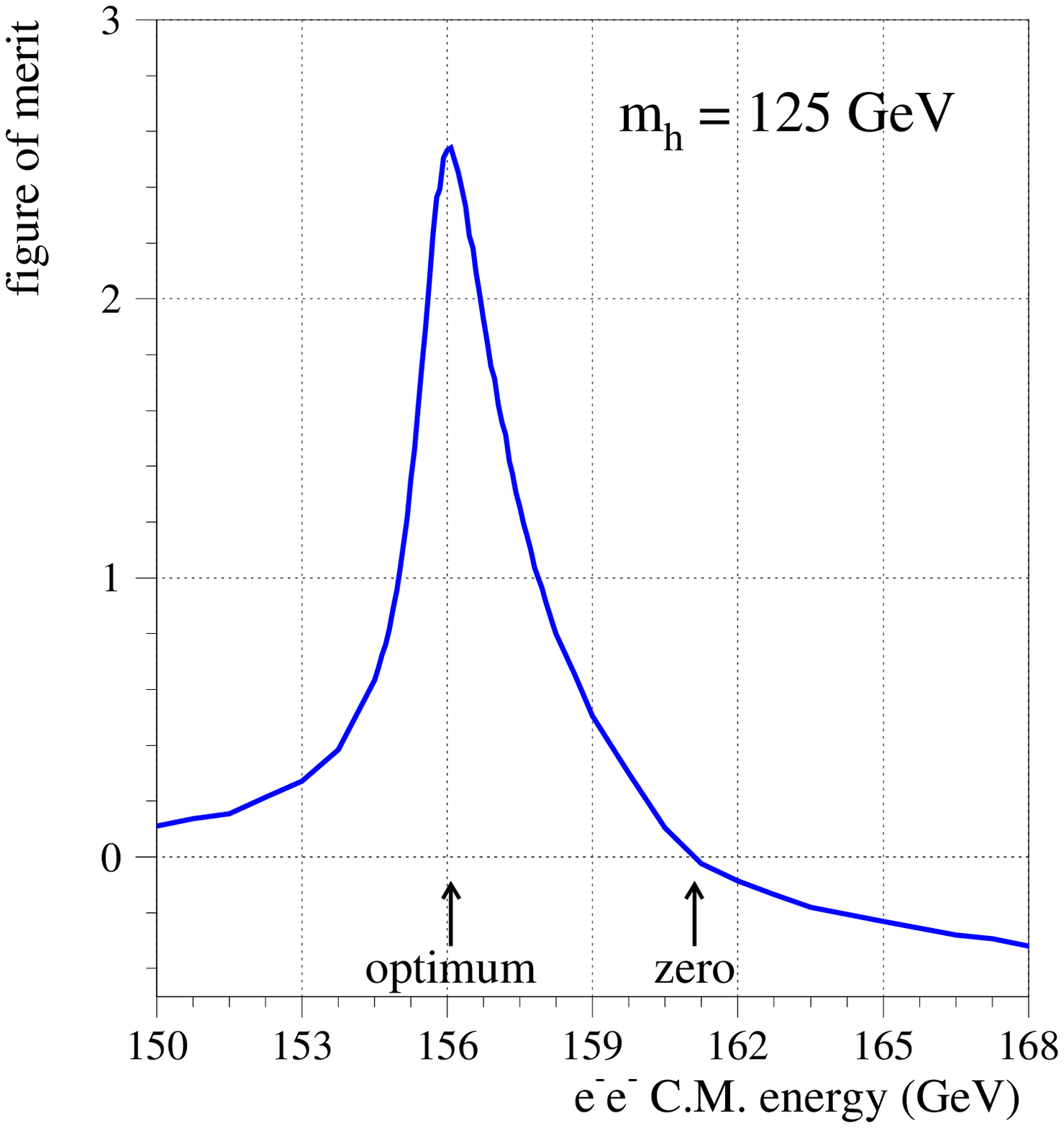,height=7cm}
\epsfig{file= 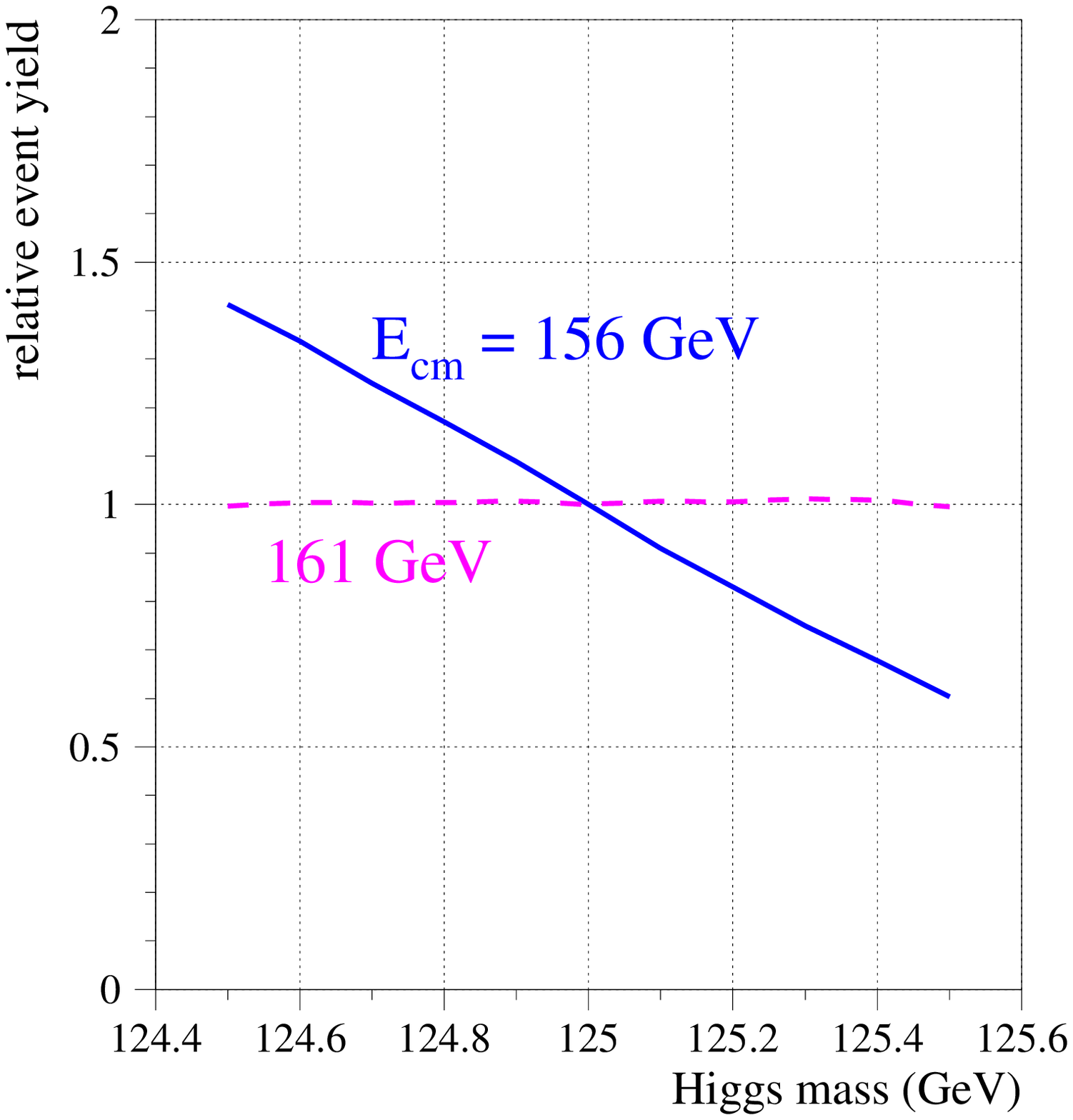,height=7cm}
\epsfig{file= 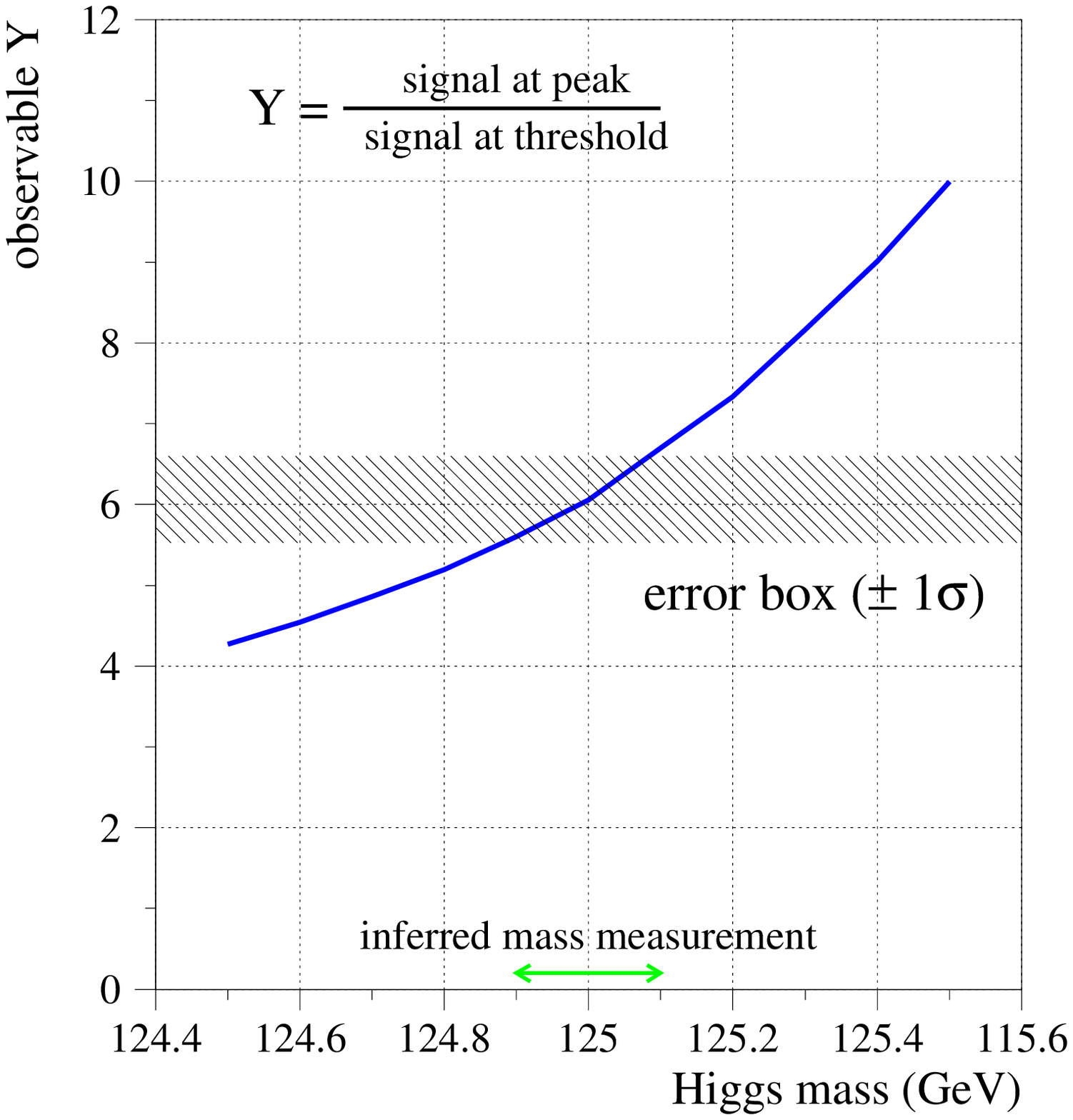,height=7cm}}
\end{center}
\caption[.]{\label{fig:scan}
\em 
(a) A figure of merit quantifying the measurement error on $M_h$ as a function of the $E_{CM} (e^-e^-)$. 
The energies with maximum and zero sensitivity are marked.
(b) The relative yields of a 125~GeV Higgs boson at the points of maximum and zero sensitivity to $M_h$.  
(c) Behavior of the observable $Y \equiv$ signal at peak/signal at threshold as a function of $m_H$, 
and the projected error.
}
\end{figure}

\noindent
\underline{Mass Measurement}\\
The sharp edge of the $\gamma \gamma$ luminosity function seen in 
Fig.~\ref{fig:spectra} and \ref{fig:sapphireSpectra}  provides an opportunity to measure $M_h$ accurately by
varying the electron beam energy. As it crosses the Higgs production threshold, 
the number of Higgs events increases dramatically, as reflected in the sharp excitation
curves in Fig.~\ref{fig:excitation}. The position of this rise enables one to measure the Higgs mass,
as discussed in~\cite{Ohgaki} and in the context of CLICHE in~\cite{asner}, where it was shown that
the point of maximum sensitivity to the Higgs mass is a few GeV below the peak of the cross section.
It was found that a run of one year at the peak and half a year each on and below threshold
would enable $M_h$ to be measured with an accuracy $\sim 100$~MeV, see Fig.~\ref{fig:scan}.

\noindent
\underline{$h \to {\bar b} b$ Decay}\\
The large branching ratio for $h \to {\bar b} b$ decay for $M_h \sim 125$~GeV makes it
the main channel for Higgs studies at a $\gamma \gamma$ collider, so this channel has 
received considerable attention. We analyzed this in the context of the CLICHE study~\cite{asner},
including perturbative QCD backgrounds, assuming cuts with an efficiency of 85 to 90\% for the 33\% 
of ${\bar b} b$ events that do not contain neutrinos, 70\% efficiency for double-tagging of
${\bar b} b$ final states and a 3.5\% ${\bar c} c$ contamination. Results from the CLICHE study
for $M_h = 115$~GeV are shown in Ref.~\cite{asner}, and we expect similar results for $M_h \sim 125$~GeV.
The corresponding accuracy in measuring $h \to {\bar b} b$ decay is 2\%, as reported in Table~\ref{table:decay}.

\begin{figure}[htbp]
\begin{center}
\mbox{\epsfig{file=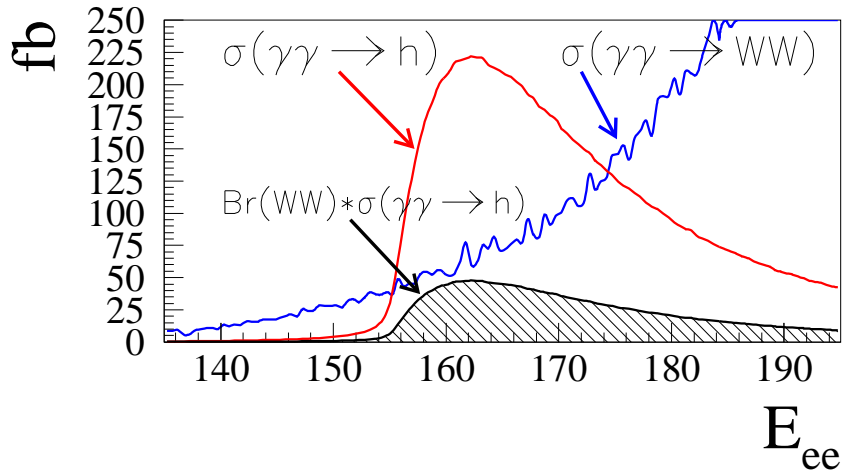,height=8cm}}
\end{center}
\caption[.]{\label{fig:wwcross}      
\em
 Cross sections for $\gamma\gamma\rightarrow h$, 
$\gamma\gamma\rightarrow h \times {\cal BR}(h\to WW)$ for 
$m_H=125$~GeV and  $\gamma\gamma\rightarrow WW$
production, as calculated using {\tt Pandora}~\cite{Pandora}. 
}
\end{figure}

\noindent
\underline{$h \to W W^*$ Decay}\\
This decay was also studied in~\cite{asner} for $M_h = 115$~GeV. In the case, the
situation for $M_h \sim 125$~GeV is somewhat different, since on the one hand the
expected $h \to W W^*$ branching ratio is higher than for $M_h = 115$~GeV, but on
the other hand the cross section for the background process $\gamma \gamma \to W W(^*)$
increases rapidly with the centre-of-mass energy~\cite{Pandora}, see Fig.~\ref{fig:wwcross}. 
In the absence of a more detailed
study, however, we expect that a precision similar to that found in~\cite{asner}
could be achieved, namely $\sim 5$\% as also reported in Table~\ref{table:decay}.

\noindent
\underline{$h \to \gamma \gamma$ Decay}\\
The decay  $h \to \gamma \gamma$ is rare one, but 
the large number of Higgs events a $\gamma \gamma$ collider would include an interesting number of
$h \to \gamma \gamma$ events. Moreover, the backgrounds are expected to be small, and
initial estimates in~\cite{asner} indicated that a peak should
be observable in the $\gamma \gamma$ mass distribution.
The quadratic dependence of the number of events $\sim \Gamma^2_{\gamma \gamma}/\Gamma_{total}$
implies that if $\Gamma_{total}$ could be measured accurately elsewhere, a $\gamma \gamma$
collider would yield a small error in $\Gamma_{\gamma \gamma}$. Conversely, if $\Gamma_{\gamma \gamma}$
were to be measured elsewhere, a small error $\Gamma_{total}$ could be obtained.
In Table~\ref{table:decay} we see that a 10\% measurement or better of $\sim \Gamma^2_{\gamma \gamma}/\Gamma_{total}$
could be made within a year of data-taking.

\noindent
\underline{Other Decay Modes}\\
We expect that $h \to Z Z^*$ and  $h \to Z \gamma$ should be observable at a $\gamma \gamma$ collider, 
but have not made any studies. Pessimism was expressed in~\cite{asner} about the prospects
for observing $h \to \tau^+ \tau^-$, but we think this should be revisited. Likewise, Ref.~\cite{asner}
was pessimistic about measuring $h \to {\bar c} c$ and $gg$, in these case probably correctly.

\noindent
\underline{Summary of Possible Measurements}\\
We have summarized the possibilities for measurements in the $h \to {\bar b}b, WW^*$
and $\gamma \gamma$ channels. The observabilities and plausible statistical errors in
measuring the products $\sigma(\gamma \gamma \to h) {\cal BR}(h \to X)$ for these
channels are reported in Table~\ref{table:decay}. Studies conducted in the context
of CLICHE indicated that the systematic errors could be controlled to similar levels. In addition,
$M_h$ could be measured in four ways (fitting the peaks in the ${\bar b} b$, $\gamma \gamma$ and $ZZ^*$
mass distributions, and by the threshold method). Moreover, possible CP asymmetries could be measured 
with a precision of about 5\%.

\begin{table}[t]
\caption{\it The statistical errors on selected decay modes of
a 125~GeV Higgs boson in the Standard Model, calculated for 
a sample of 20,000  Higgs bosons corresponding to one year with the nominal luminosity of
CLICHE or SAPPHiRE.}
\label{table:decay}
\begin{center}
\begin{tabular}{lc||c|c|c|c}
\hline
decay mode\,\,\,\,\,  &\,\,\,\,\, raw events/year\,\,\,\,\, &\,\,\,\,\, S/B\,\,\,\,\,  & \,\,\,\,\,\,\,\,\,\,$\epsilon_{sel}$\,\,\,\,\, &\,\,\,\,\,\,\,\,\,\, ${\cal BR}$\,\,\,\,\, &\,\,\,\,\, $\Delta \Gamma_{\gamma\gamma} {\cal BR}/\Gamma_{\gamma\gamma}{\cal BR}$ \\
\hline\hline
${\bar b} b$      & 11540         & 4.5  &   0.30 &  57.7\% & 2\% \\
\hline
$W^+ W^-$         & 4300            & 1.3  &  0.29  & 21.5\% & 5\% \\
\hline
$\gamma\gamma$    &  45             & ---  &  0.70  &  0.23\% & 8\% \\
\hline\\
\end{tabular}
\end{center}
\end{table}

\section{Final Comments}

A $\gamma \gamma$ collider could produce a number of Higgs bosons comparable
to an $e^+ e^-$ Higgs factory. The backgrounds from
conventional $\gamma \gamma$ collisions would be larger than the $e^+ e^-$-induced backgrounds to the reaction
$e^+ e^- \to Z h$, and an $e^+ e^-$ collider would have other possibilities at
other energies, e.g., for $Z$ studies at lower energies, or for ${\bar t} t$ studies at
higher energies. Nevertheless, we feel that the Higgs physics programme of a $\gamma \gamma$
collider is of comparable interest to that with an $e^+ e^-$ collider, bearing in mind, e.g., the
possibilities for CP studies that we have not discussed here.

In this note we have presented two concepts for a $\gamma \gamma$ collider: the
CLICHE idea based on CLIC~1~\cite{asner}, and the SAPPHiRE idea based on the recirculating linacs
envisaged for the LHeC~\cite{lheccdr}. These concepts therefore offer considerable synergies with these
other projects. Moreover, we note two generic advantages of $\gamma \gamma$ colliders
over $e^+ e^-$ colliders: they need a lower centre-of-mass energy and they do not need a 
positron source. Both of these features offer potential economies, though they may
be offset by other disadvantages, such as the need for a high-performance laser
backscattering system. This is presumably the aspect of a $\gamma \gamma$ collider
that requires the most R\&D, though one should be able to piggy-back on the developments
in laser systems made for other purposes.

Finally, we note that, although the SAPPHiRE concept was motivated by the
recirculating linac system proposed for the LHeC, it is not limited to the context
of that project. One could well imagine building such a recirculating linac system
independently, and it might provide an appealing, timely and cost-efficient stand-alone
possibility for a Higgs factory.

\section*{Acknowledgements}

This work of JE was supported partly by the London
Centre for Terauniverse Studies (LCTS), using funding from the European
Research Council via the Advanced Investigator Grant 267352.


\begin{thebibliography}{99}
%
\bibitem{:2012gu}
  G.~Aad {\it et al.}  [ATLAS Collaboration],
{\it Observation of a new particle in the search for the Standard Model Higgs boson with the ATLAS detector at the LHC,}
  arXiv:1207.7214 [hep-ex];
S.~Chatrchyan {\it et al.}  [CMS Collaboration],
{\it Observation of a new boson at a mass of 125 GeV with the CMS experiment at the LHC,}
  [arXiv:1207.7235 [hep-ex]].
%
%

\bibitem{CLIC}
E.~Accomando {\it et al.}  [CLIC Physics Working Group],
{\it Physics at the CLIC multi-TeV linear collider}, CERN Yellow Report CERN-2004-005,
  hep-ph/0412251.

\bibitem{asner} 
D.~Asner {\it et al.},
{\it Higgs physics with a gamma gamma collider based on CLIC I},
  Eur.\ Phys.\ J.\ C {\bf 28} (2003) 27
  [hep-ex/0111056].

\bibitem{lheccdr} J. Abelleira Fernandez et al, 
{\it A Large Hadron Electron Collider at CERN - Report on the Physics and Design Concepts for Machine and Detector},   
{\it Journal of Physics G: Nuclear and Particle Physics} {\bf 39} Number 7 (2012) arXiv:1206.2913 [physics.acc-ph].

\bibitem{laseripac12}
A.~T\"{u}nnemann, T.~Eidam and J.~Limpert, 
{\it Advanced Solid-State Lasers are Merging with Accelerators},
Proc.~IPAC~2012, New Orleans (2012).

\bibitem{dimad}
R.~Servranckx, K.~L.~Brown, L.~Schachinger and D.~Douglas,
{\em  SLAC-0285}; \\
{\tt
 http://www-project.slac.stanford.edu/lc/local/AccelPhysics/Codes/Dimad/}.

\bibitem{cain2} P.~Chen, G.~Horton-Smith, T.~Ohgaki, A.~W.~Weidemann 
and K.~Yokoya, {\em Nucl. Instrum. Meth.} {\bf A355}, 107 (1995); \\
See
{\tt http://www-acc-theory.kek.jp/members/cain/cain21b.manual/main.html}.

\bibitem{bogacz} 
S.A. Bogacz, I. Shin, D. Schulte and F. Zimmermann, 
{\it LHeC ERL Design and Beam-Dynamics Issues}, 
Proc. IPAC~2012, New Orleans (2012), p. 1120.

\bibitem{GP}
D.~Schulte, {\tt http://www-project.slac.stanford.edu/lc/bdir/programs/guinea\_pig/}
{\tt gp\_index.html}.

\bibitem{teng} 
L. Teng, {\it Minimizing the Emittance in Designing the Lattice of an Electron Storage Ring}, FNAL TM-1269 (1984).

\bibitem{brinkmann}
 R. Brinkmann, Ya. Derbenev and K. Fl\"{o}ttmann, 
{\it A Flat Beam Electron Source for Linear Colliders}, TESLA Note 99-09 (1999).

\bibitem{piot} 
P. Piot, Y.-E. Sun and K.-J. Kim, 
{\it Photoinjector Production of a Flat Beam with Transverse Emittance Ratio of 100}, Proc.~LINAC~2006 Knoxville (2006).

\bibitem{Ohgaki}
T.~Ohgaki, {\em  Int. J. Mod. Phys.} {\bf  A15} (2000) 2605.

\bibitem{Pandora}
M.~E.~Peskin,
{\it Pandora: An Object oriented event generator for linear collider physics,}
  hep-ph/9910519;
{\tt http://www-sldnt.slac.stanford.edu/nld/new/docs/generators/pandora.htm}.

\end{thebibliography}
\end{document}